\documentstyle[psfig,emulateapj]{article}

\lefthead{Barger et al.}

\begin{document}
\title{Redshift Distribution of the Faint Submillimeter Galaxy Population}
\author{A.\,J.\ Barger,$\!$\altaffilmark{1,6}
L.\,L.\ Cowie,$\!$\altaffilmark{1,6}
I.\ Smail,\altaffilmark{2} R.\,J.\ Ivison,\altaffilmark{3}
A.\,W.\ Blain,\altaffilmark{4} J.-P.\ Kneib\altaffilmark{5}
}

\altaffiltext{1}{Institute for Astronomy, University of Hawaii,
2680 Woodlawn Drive, Honolulu, Hawaii 96822, USA}
\altaffiltext{2}{Department of Physics, University of Durham, South Road,
Durham, DH1 3LE, UK}
\altaffiltext{3}{Institute for Astronomy, Department of Physics \& Astronomy,
University of Edinburgh, Blackford Hill, Edinburgh, EH9 3HJ, UK}
\altaffiltext{4}{Cavendish Laboratory, Madingley Road, Cambridge CB3 OHE, UK}
\altaffiltext{5}{Observatoire Midi-Pyr{\'e}n{\'e}es, 14 Avenue E.~Belin,
F-31400 Toulouse, France}
\altaffiltext{6}{Visiting Astronomer, W.\,M.\ Keck Observatory, jointly
operated by the California Institute of Technology and the University
of California}

\slugcomment{Accepted by the Astronomical Journal for the June 1999 issue}

\begin{abstract}
We present a Keck\,II LRIS spectroscopic follow-up study of the possible 
optical counterparts to a flux-limited sample of galaxies selected from 
an 850-$\mu$m survey of massive lensing clusters using the SCUBA 
bolometer array on the JCMT.
These sources represent a population of luminous dusty galaxies
responsible for the bulk of the 850-$\mu$m background detected by {\it
COBE} and thus for a substantial fraction of the total far-infrared
emission in the Universe.  We present reliable redshifts for 20
galaxies and redshift limits for a further four galaxies selected from
the error-boxes of 14 submillimeter (submm) sources.  Two other
submm detections in the sample have no obvious optical counterparts, and
the final submm source was only identified from imaging data after the 
completion of our spectroscopic observations.  The optical identifications 
for 4 of the submm sources
have been confirmed through either their detection in CO at
mm-wavelengths (two pairs of galaxies at $z=2.55$ and $z=2.80$) or from
the characteristics of their spectral energy distributions  (two of the
central cD galaxies in the lensing clusters).  Plausible arguments
based on the optical spectral properties (starburst or AGN signatures)
of the counterparts allow us to identify a further two likely
counterparts at $z=1.06$ and 1.16.  For the remaining 8 cases, it is
not always clear which, if any, of the optical sources identified are
the true counterparts.  Possible counterparts for these have redshifts
ranging from $z=0.18$ to $z=2.11$. The application of a range of
techniques, including near- and mid-infrared imaging and radio mapping,
will assist in the identification of the true sources of the submm
emission, while CO line mapping with current mm-interferometers and
hard X-ray observations should aid in the determination of the nature of
their emission.  Working with the current identifications, we suggest
that the majority of the extragalactic background light in the submm is
emitted by sources at $z<3$ and hence that the peak activity in
highly-obscured sources (both AGN and starbursts) lies at relatively
modest redshifts.  We find that a lower limit of 20\ per cent of the
submm sources in our sample show some sign of AGN activity;
however, we caution that this does not necessarily translate into a
20\ per cent AGN contribution to the measured submm emission from these sources.
\end{abstract}


\section{Introduction}\label{intro}

%
% Figure 1
%

\begin{figure*}[tbh]
\centerline{\psfig{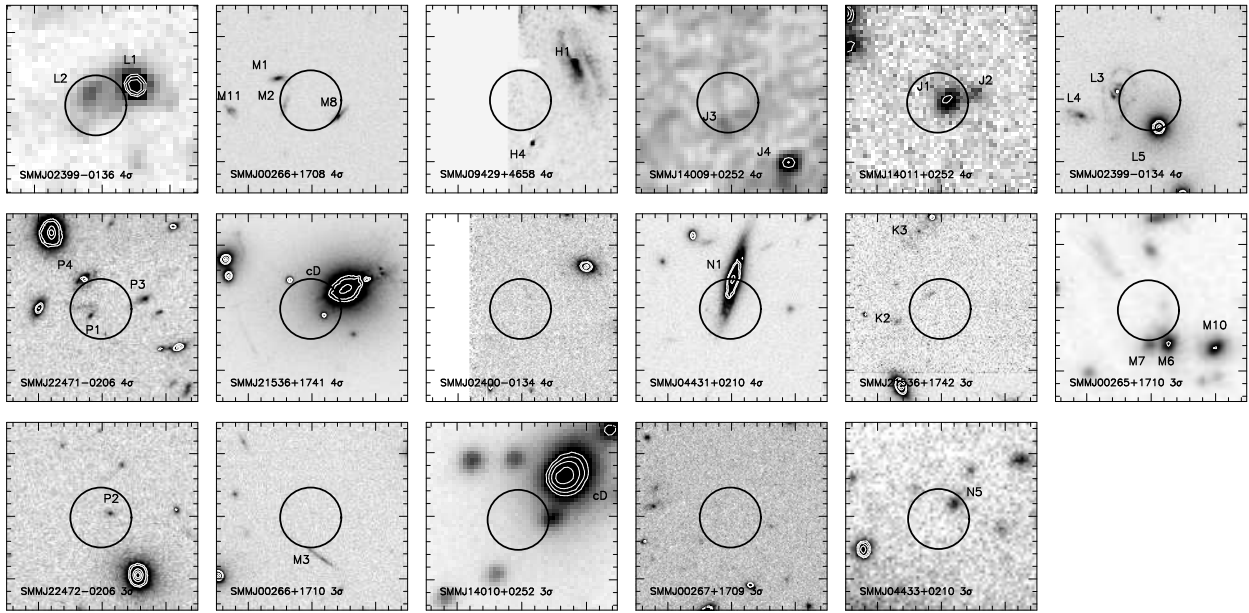}}
\figurenum{1}
\figcaption[barger.fig1.ps]{
%{\sc Fig.}~1.---
$15''\times 15''$ images of the 16 submm sources from
the Smail et al.\ (1998) cluster survey that have corresponding optical
imaging. The orientations of the images are arbitrary.
The objects are ordered from the upper-left by decreasing
apparent 850-$\mu$m flux. Error circles of 3$''$ radius are centered on
the submm source positions. The possible optical counterparts in each field
are labeled and can be cross-referenced with the data in Table~2.
\label{fig1}}
\end{figure*}

The cumulative emission from all objects lying beyond the Galaxy, the
extragalactic background light (EBL), provides important constraints on
the integrated star formation history of the Universe. The recent
measurement of the EBL at far-infrared (FIR) and submm wavelengths
using data from the {\it FIRAS} and {\it DIRBE} experiments on the {\it
COBE} satellite (\markcite{puget96}Puget et al.\ 1996;
\markcite{guider97}Guiderdoni et al.\ 1997;
\markcite{schlegel97}Schlegel et al.\ 1998; \markcite{fixsen98}Fixsen
et al.\ 1998; \markcite{hauser98}Hauser et al.\ 1998) indicates that
the total emission from star formation and AGN activity that is
absorbed by dust and reradiated into the FIR/submm is 
comparable to the unobscured emission seen in the optical. This
suggests that obscured star formation may be responsible for a large
fraction of the stars and metals seen in the local Universe, a
conclusion which would have profound consequences for models of galaxy
formation and evolution.

Independent support for the claim that optical estimates of the star
formation density in the distant Universe may be missing a substantial
component that is obscured by dust comes from deep submm surveys with
the new camera SCUBA (Submillimeter Common User Bolometer Array;
\markcite{holland98}Holland et al.\ 1999) on the 15-m James Clerk Maxwell
Telescope\footnote{The JCMT is operated by the Joint Astronomy Center 
on behalf of the parent organizations, the Particle Physics and Astronomy 
Research Council in the United Kingdom, the National Research Council of 
Canada, and the Netherlands Organization for Scientific Research.}
(JCMT) on Mauna Kea. SCUBA has, for the
first time, enabled deep, unbiased surveys to be made of the submm sky
(\markcite{smail97}Smail, Ivison \& Blain 1997;
\markcite{barger98}Barger et al.\ 1998, 1999; \markcite{hughes98}Hughes et
al.\ 1998; \markcite{eales99}Eales et al.\ 1999). These surveys have
uncovered numerous sources with
properties similar to those expected for distant ultraluminous
infrared galaxies (\markcite{barger98}Barger et al.\ 1998;
\markcite{smail98}Smail et al.\ 1998). If the majority of the submm
emission in these systems comes from dust-obscured star formation, then
their inferred star formation rates are of the order of several hundred
solar masses per year.

The deepest submm counts (\markcite{blain99b}Blain et al.\ 1999b),
fluctuation analyses (e.g.\ \markcite{hughes98}Hughes et al.\ 1998), and
analyses of deep and wide-area surveys
(e.g.\ \markcite{barger99}Barger et al.\ 1999) 
indicate that the bulk of the background emission at 850-$\mu$m detected 
by {\it COBE} is resolved into discrete sources at a flux limit of 
$<1$\,mJy. Thus, we are in a
position to undertake detailed studies of the population responsible
for the majority of the emission in the FIR background. In particular,
we can measure the redshift distribution of the submm population and
use this to trace the extent and evolution of obscured star formation
in the distant Universe.  The detailed study of the resolved component of
the background also provides a clear method for determining
what fraction of the submm emission originates from 
AGN activity rather than star formation 
(\markcite{blain99a}Blain et al.\ 1999a;
\markcite{guider98}Guiderdoni et al.\ 1998).

Finally, it is conceivable that some fraction of the population
responsible for the FIR background lies at high redshift, $z\sim 5$--10.
Submm selection is a powerful technique for locating such distant
galaxies.  Indeed, the steep thermal dust spectrum, which peaks in the
FIR at a rest-frame wavelength of about 100-$\mu$m, is redshifted into the
submm for $z>1$.  The resulting strong negative K-correction
for sources out to $z\sim 10$ is sufficient to offset cosmological dimming
for $q_0=0.5$.  Even for low values of $q_0$ the 850-$\mu$m flux density
is only expected to decrease by a factor of a few over this redshift
range (\markcite{bl93}Blain \& Longair 1993; \markcite{hughes97}Hughes,
Dunlop \& Rawlings 1997).

In this paper we present spectroscopic redshift information for a
sample of submm-selected galaxies from the SCUBA cluster lens survey of
\markcite{smail97}Smail et al.\ (1997, 1998). One advantage of a 
lensed survey is that the problem of source confusion (which, due to the 
coarse resolution of submm telescopes, can contribute noise in faint images;
\markcite{blain98}Blain, Ivison, \& Smail 1998) is reduced
since both the flux densities and the mean separations on the sky of the 
background sources are increased
(\markcite{blain99}Blain et al.\ 1999b). The primary advantage,
however, is that the clusters magnify any background sources (here 
the median amplification of the submm fluxes is $\sim 2.5$), thereby
providing otherwise unachievable sensitivity in the submm and easing
spectroscopic follow-up in the optical. The full 850-$\mu$m survey
detected 17 sources above $3\sigma$ significance ($1\sigma<2$\ mJy in
the image plane) over a total surveyed area of 36\,arcmin$^2$. Crude
redshift limits for these sources were derived from broad-band imaging
by \markcite{smail98}Smail et al.\ (1998), who inferred that most of
the galaxies were at $z\leq 5$.

Section 2 describes the submm sample and the spectroscopic follow-up.
Section 3 presents the redshift identification for each submm source.
Finally, section 4 discuss the results and gives our conclusions.

%
% TABLE 1
%
{\scriptsize
\begin{deluxetable}{lccc}
\tablewidth{200pt}
\tablenum{1}
\label{table-1}
\tablecaption{Log of the Keck\,II LRIS spectroscopic observations\label{tab1}}
\tablehead{
\colhead{UT Date} & \colhead{Seeing} & \colhead{Targets} &
\colhead{Exposure} \\
\colhead{} & \colhead{($''$)} & \colhead{} & \colhead{(ks)}
}
\startdata

1998 July 18 & 0.6~ & J1/J2, J3   & $3.6$ \\
             &      & K2, K3      & $3.6$ \\
             &      & P1, P2      & $3.6$ \\
             &      & M4, M6, M10 & $1.8$ \\
             &      & M1, M2      & $2.4$ \\
1998 Aug 21  & 0.6~ & P1, P4      & $3.6$ \\
             &      & L3, L4      & $3.6$ \\
1998 Aug 22  & 0.6~ & P1, P2      & $3.6$ \\
             &      & M1, M3, M11 & $2.4$ \\
1998 Sept 16 & 0.6~ & P1, P2      & $3.1$ \\
             &      & M1, M3      & $3.6$ \\
1998 Sept 17 & 0.6~ & K2, cD      & $3.6$ \\
             &      & M1, M3      & $3.6$ \\
             &      & N1, N2      & $3.6$ \\
1998 Oct 22  & 0.6~ & H4          & $3.0$ \\
\enddata
\end{deluxetable}}

\section{Sample and Observations}\label{observ}

\markcite{smail98}Smail et al.\ (1998) sought the optical counterparts
for their 17 submm detections within a conservative $\lesssim
6$\,arcsec radius (the combined $2\sigma$ positional uncertainty of
their faintest sources) from the nominal submm position. All
possible optical counterparts were identified in deep {\it HST}
and ground-based imaging data to $I\sim 26.0$ and 23.5, respectively,
for the 16 submm sources covered by existing optical imaging
data. Using a deep Keck\,II LRIS $I$-band image, we recently identified
a probable optical counterpart ($I=24.3$) to the seventeenth source, 
SMM\,J0443+0210 (N5; Fig.~\ref{fig1}). Unfortunately, this identification
came too late for the source to be included in the spectroscopic
campaign described here.  Two of the 17 sources have no obvious
counterparts to the magnitude limits of their respective images.  

Since the lensing clusters lie outside the Galactic plane, the 
850-$\mu$m sources are unlikely to arise from local regions, and
indeed we find no evidence for dark patches caused by dust foreground
globules in any of the optical images.
Furthermore, the weak or non-detection of these sources at 450-$\mu$m
(\markcite{smail97}Smail et al.\ 1997) rules out their origins being
local.

As noted by Smail et al.\ (1998), since the optical counterparts to the
submm sources are likely to be faint and the cluster fields are
crowded, there is some likelihood that unrelated galaxies will fall 
within the submm error boxes.
The probability of a galaxy with a given observed apparent magnitude
$m$ falling at random within a circle of radius $r$ centered on the
submm source is proportional to $r^2$ times the number density per 
surface area on the sky of galaxies with magnitude $m$.
Smail et al.\ estimated on an object-by-object basis 
(from the observed galaxy number counts for each optical image)
the probability that a galaxy with the observed apparent magnitude or brighter 
would fall at random within a circle defined by the optical and submm positions.
Although this procedure provides some measure of the reliability of the 
identifications, it does not allow the definitive determination of
the true counterparts to individual submm sources. Thus, we have simply
used these probability estimates to determine the priority with
which to target sources, but in most cases
we have targeted all visible counterparts within the error circles. 

The optical spectroscopy of the candidate submm sources was undertaken
with the Low-Resolution Imaging Spectrometer (LRIS; \markcite{oke95}Oke
et al.\ 1995) on the Keck\,II 10-m telescope using a wide 1.5\,arcsec
long-slit during several runs from 1998 July to 1998 October. With the
wide slit and the 300 lines\,mm$^{-1}$ grating blazed at 5000\,\AA, the
resolution was 14\,\AA. The wavelength coverage was varied slightly,
depending on the target and whether there were additional emission
lines to search for. Two objects were observed per slit, which defined
the position angle. The observations were typically 1\,hr per slit,
broken into three sets of exposures. Some of the objects were re-observed
several times to improve the signal-to-noise. Table~1 
gives a log of the
Keck spectroscopic observations, including date, seeing, targets, and
exposure times. The objects were stepped along the slit by 10\,arcsec in each
direction, and the sky backgrounds were removed using the median of the images
to avoid the difficult and time-consuming problems of flat-fielding LRIS
data. Details of the spectroscopic reduction procedures can be found in
\markcite{cowie96}Cowie et al.\ (1996).

Figure~\ref{fig1} shows the fields for the 17 submm sources from the SCUBA
cluster lens survey, including the newly identified source SMM\,J0443+0210.
The individual images are centered on the submm positions with nominal
3\,arcsec (1\,$\sigma$) radius error circles marked. 
Each square box covers an area $15''\times 15''$;
the orientations of the images are arbitrary. 
The possible optical counterparts for
each of the submm detections are labeled by name and can be
cross-referenced with Table~\ref{table-2}. In both Fig.~\ref{fig1} and 
Table~\ref{table-2} the submm
sources are ordered by apparent 850-$\mu$m flux for the $4\sigma$
and $3\sigma$ samples.
The columns in Table~\ref{table-2} include the submm source name, apparent
850-$\mu$m flux, candidate optical counterparts
(most likely listed first), $I$ magnitudes for the counterparts, redshifts
for the counterparts, name and redshift of the cluster field, and lensing 
amplifications for the counterparts at the given redshifts.

%
% Figure 2
%

\begin{figure*}[tbh]
\figurenum{2}
\centerline{\psfig{figure=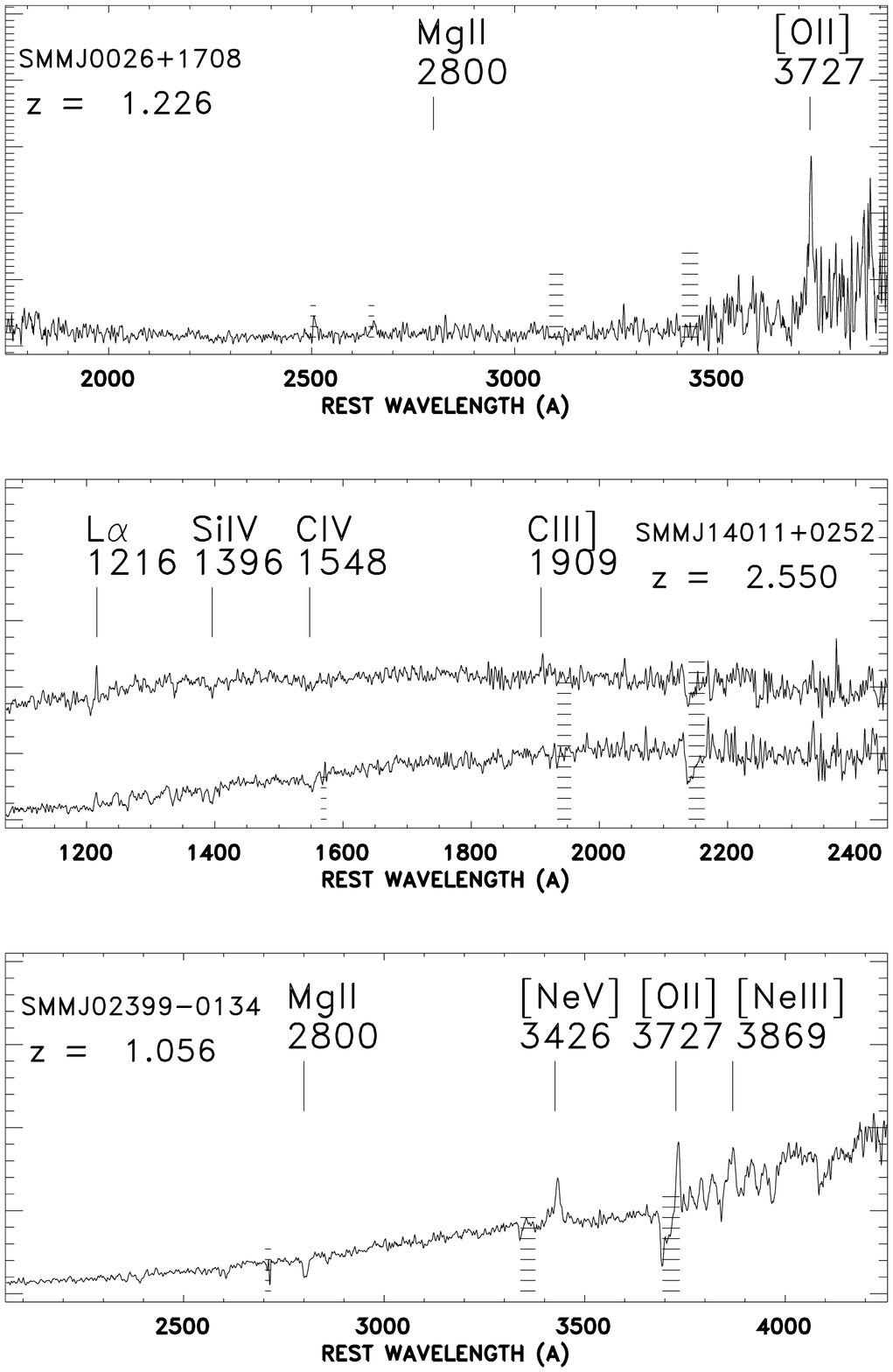,width=3.1in}
\psfig{figure=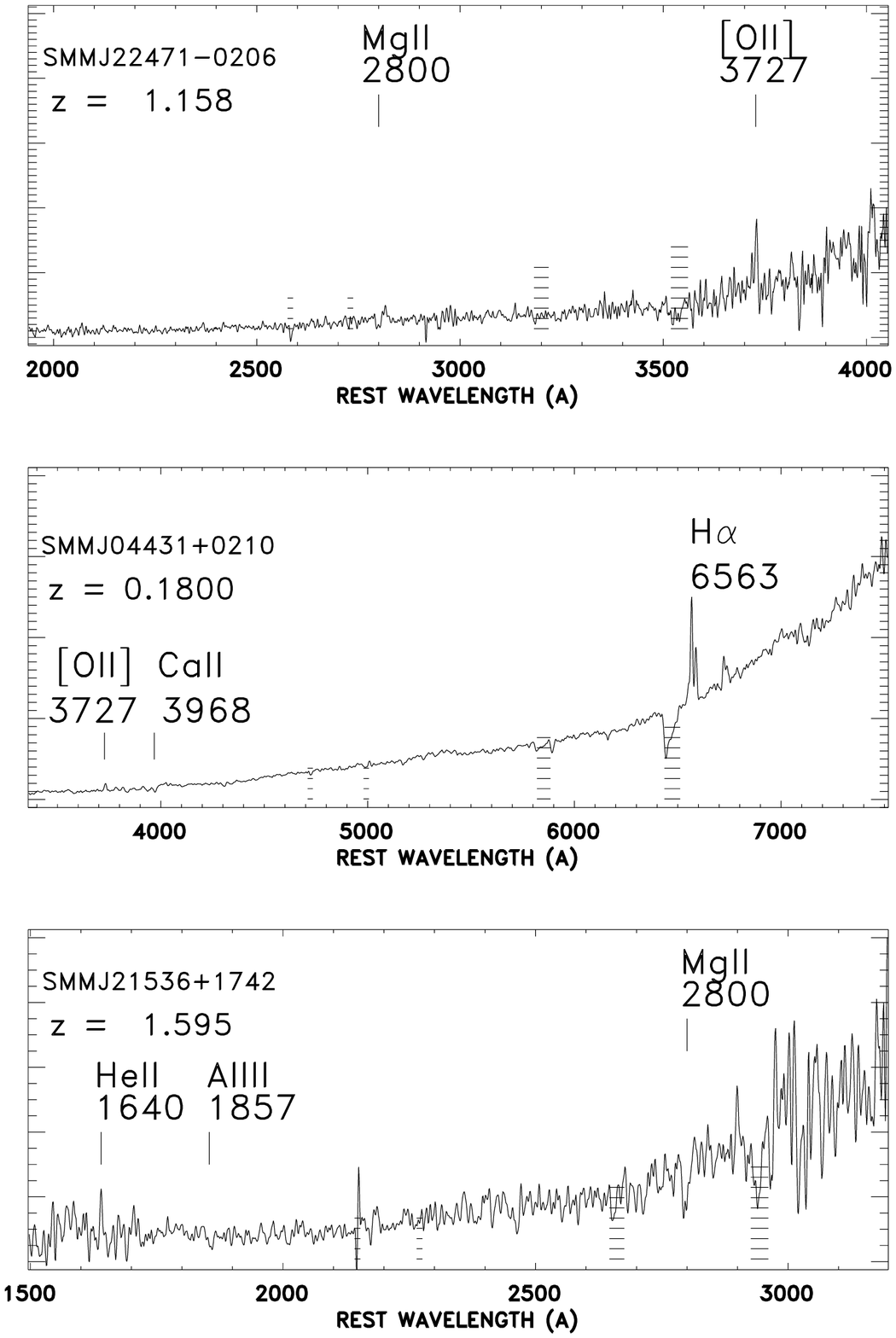,width=3.1in}}
\centerline{\psfig{figure=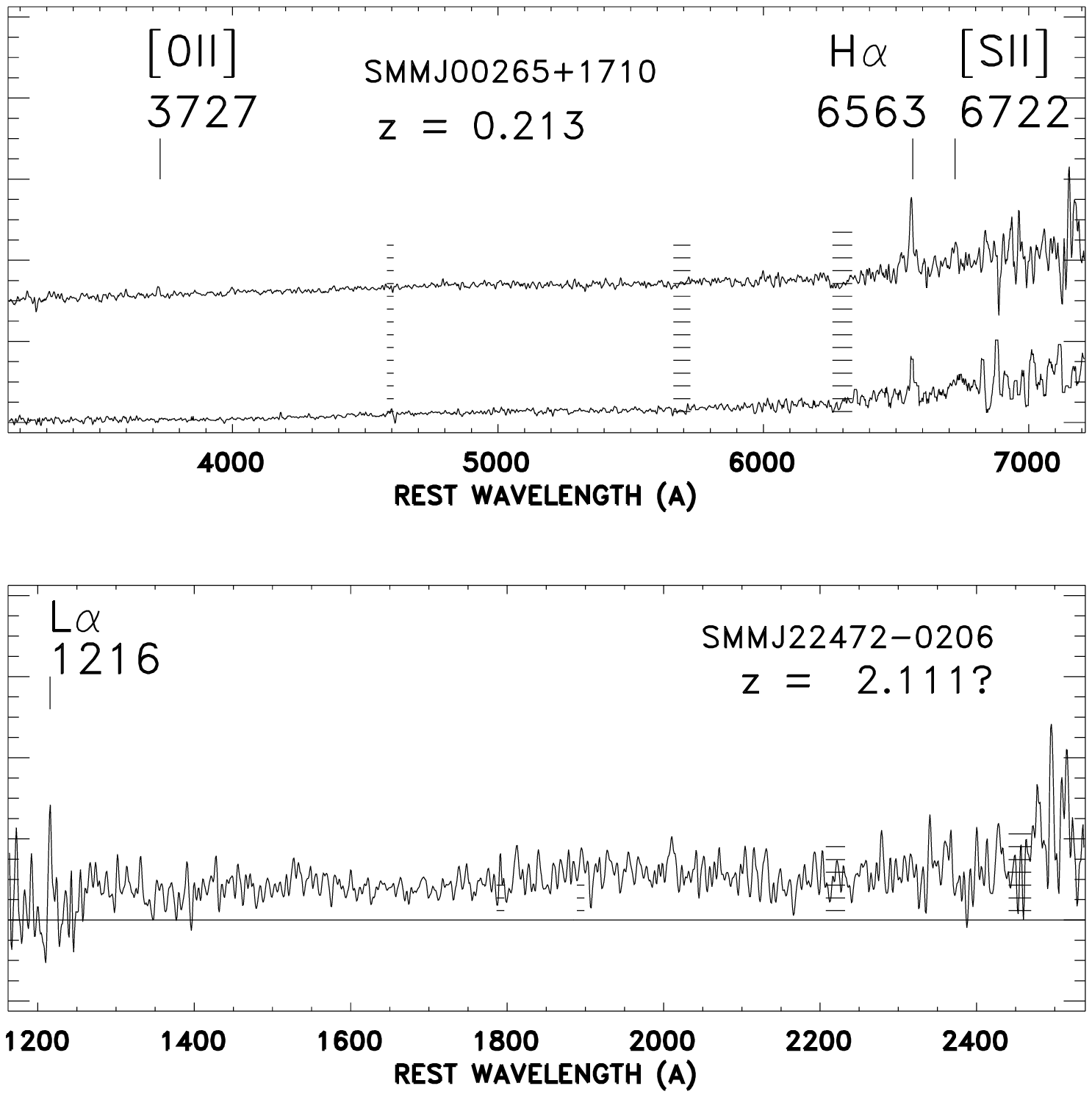,width=3.1in}}
\vspace*{-1.0in}
\figcaption[barger.fig2abc.ps, barger.fig2def.ps, barger.fig2gh.ps]{
%{\sc Fig.}~2.--- 
Keck spectra, where available, of the most likely optical 
counterparts to the submm detections:
(a)\ M2 ($z=1.23$), (b)\ J1/J2 ($z=2.55$), (c)\ L3 ($z=1.06$),
(d)\ P4 ($z=1.16$), (e)\ N1 ($z=0.18$), (f)\ K2 ($z=1.60?$),
(g)\ M6/M10 ($z=0.21$), and (h)\ P2 ($z=2.11?$).
For (b)\ J1/J2 and (g)\ M6/M10 the spectra of both members of the galaxy
pairs are shown. The spectrum for J3 is not shown because the redshift
is inconclusive. \label{fig2}}
\end{figure*}

\section{Redshift Identification}

Our ability to conclude whether a particular galaxy is likely to be
the submm source depends strongly on whether the optical spectrum of
the galaxy shows any remarkable features, such as particularly strong
[O{\sc ii}]\,$\lambda 3727$ or H\,$\alpha$ emission lines that indicate 
a starburst, or high excitation or broad lines that show the presence of an 
AGN. The relative paucity of AGNs in the general field population ($\lesssim
1$\%) suggests that if an AGN is identified, then the submm emission is
most likely associated with that source.
We note, however, that the visibility of remarkable features, especially 
AGN lines, will be strongly affected by dust in the galaxy -- the same component
which our submm selection should guarantee is present.  The identification
process is therefore somewhat problematic, and while we are capable
of robustly identifying some submm sources with optical counterparts,
specifically those with the most striking spectral features (as has
been confirmed through the CO detection of two of our candidates), this
approach leaves us with a number of ambiguous cases.  As a secondary
criterion, we consider the morphologies of the objects and the presence
of any merger activity, which can be an indicator of luminous FIR
systems in the local Universe.

We now discuss the nature and reliability of our redshift measurements in
order of decreasing submm flux density, as presented in Fig.~\ref{fig1} and 
Table~\ref{table-2}.
In Fig.~\ref{fig2} we show the Keck spectra for the most likely 
optical counterparts to the submm sources, as discussed below.

\smallskip

\noindent{\bf SMM\,J02399$-$0136:} A redshift of $z=2.80$ was measured
for the pair of optical sources, L1/L2, by \markcite{ivison98}Ivison et
al.\ (1998). L1 is a compact, dust-obscured AGN, and L2 is a companion
structure whose extended emission may result from a vigorous burst of
star formation triggered by an interaction with L1. The detection of CO
emission in the mm-waveband coincident in redshift and position with
the optical counterpart (\markcite{frayer98}Frayer et al.\ 1998) leaves
no doubt that the L1/L2 pair is the correct identification.

%
% Figure 3
%

\begin{figure*}[tbh]
\figurenum{3}
\centerline{\psfig{figure=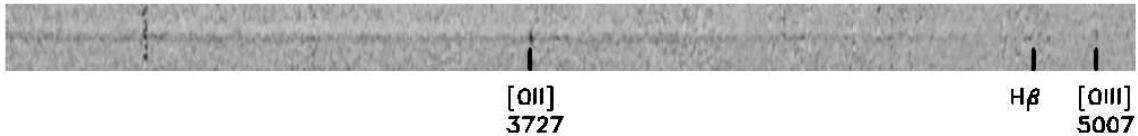,angle=90,width=6.5in}}
\vspace*{-0.1in}
\figcaption[barger.fig3.ps]{
%{\sc Fig.}~3.---
The 2D spectral image for M3 ($z=0.94$) with tick marks indicating the
positions of [O{\sc ii}]\,$\lambda 3727$, H$\beta$, and
[O{\sc iii}]\,$\lambda 5007$. Although H$\beta$ cannot be reliably identified,
the emission line identified with [O{\sc ii}] is unambiguous and that
identified with [O{\sc iii}] is also very clear.
\label{fig3}}
\end{figure*}

\noindent{\bf SMM\,J09429+4658:} The redshift for the nearby bright
galaxy H1 is $z=0.33$ (\markcite{dressler99}Dressler et al.\ 1999).  It
shows no [O{\sc ii}]\,$\lambda 3727$ or H\,$\alpha$ emission 
but contains an obvious
dust lane and thus might plausibly be the submm source.
H4 shows H\,$\alpha$ and Na\,D in absorption and is a blue star.
Another plausible counterpart --- a faint red source close to the nominal
submm position --- has recently been uncovered in
deep near-infrared imaging of the SCUBA fields
(\markcite{smail99}Smail et al.\ 1999).
Given the dust characteristics of H1 and the very low
probability of seeing such a bright foreground galaxy close to the
submm position, we consider H1 to be the most likely counterpart;
however, we caution that there is some uncertainty about this identification.

\noindent{\bf SMM\,J14009+0252:} We targeted J3, the nearest galaxy
to the nominal submm position, and found
the spectrum to be featureless and flat, so no redshift identification
could be made; however, from its blue SED and lack of obvious emission
lines, the object does not appear to lie at $z\ll 1.5$ or $z\gg 2.5$.
No spectroscopic observations were made of J4.

\noindent{\bf SMM\,J14011+0252:} We identify this submm source with the
high redshift starburst pair J1/J2 at $z=2.55$, based on strong
Ly\,$\alpha$ emission and both C\,{\sc iv}\,$\lambda 1548$ and Si\,{\sc
iv}\,$\lambda 1396$ absorption features (Fig.~\ref{fig2}b). 
This pair of classic 
Lyman break galaxies was subsequently detected in CO at close to the 
optical redshift (\markcite{frayer99}Frayer et al.\ 1999).  As with
SMM\,J02399$-$0136, the spatial and redshift coincidence of the
CO emission confirms our identification.

\noindent{\bf SMM\,J02399$-$0134:} The galaxy L3 at a redshift of
$z=1.06$ is almost certainly the submm source. The galaxy has an
unusual ring morphology (Fig.~\ref{fig1}), and the nucleus hosts a 
Seyfert 1.5 AGN,
characterized by strong [O\,{\sc ii}]\,$\lambda 3727$, [Ne\,{\sc
v}]\,$\lambda 3426$, and [Ne\,{\sc iii}]\,$\lambda 3869$ emission, in
addition to strong Mg\,{\sc ii}\,$\lambda 2800$ absorption 
(Fig.~\ref{fig2}c). L4 is at $z=0.42$ and hence is background to the cluster.
L5 is a passive cluster elliptical at $z=0.37$ (ID \#32 in 
Mellier et al.\ 1988).

\noindent{\bf SMM\,J22471$-$0206:} P4 is a $z=1.16$ emission-line galaxy
with [O\,{\sc ii}]\,$\lambda 3727$ and Mg\,{\sc ii}\,$\lambda 2800$ 
in absorption over 
weak broad Mg\,{\sc ii} emission (Fig.~\ref{fig2}d). The highly peculiar
morphology of this source, plus its weak AGN characteristics, suggest
that it is the correct optical counterpart. Other
candidates include P1, which has a flat featureless spectrum,
probably putting it at $z \simeq 2$, and the more distant (3.3$''$)
and less probable P3, which was not observed.

\noindent{\bf SMM\,J21536+1741:} 
This source lies close to the central cD galaxy of the cluster A\,2390
($z=0.23$). Since the cD galaxy is a strong radio source, there is
little doubt that this is the correct identification. The submm
emission from this source and its relation to the overall spectral
energy distribution of the central galaxy are discussed in more detail
in \markcite{edge99}Edge et al.\ (1999).

\noindent{\bf SMM\,J02400$-$0134:} There is no obvious optical
counterpart for this source down to the deep limit of the {\it HST} image.

\noindent{\bf SMM\,J04431+0210:} The bright spiral galaxy N1 shows strong
H\,$\alpha$ and [O\,{\sc ii}]\,$\lambda 3727$ 
emission (Fig.~\ref{fig2}e) and is a cluster
member at $z=0.18$ (see also \markcite{gioia98}Gioia et al.\ 1998). 
The probability of
seeing such a bright cluster galaxy only 2.3$''$ from the nominal submm
position is very low.  However, this is another field where the deep
near-infrared imaging by \markcite{smail99}Smail et al.\ (1999) 
has uncovered a faint
red source within the submm error-box. Thus, we caution that there
is still a question mark over the identification of N1 with the submm
emission.

\noindent{\bf SMM\,J21536+1742:} K2 has a relatively uncertain redshift
identification of $z=1.60$ based on weak He\,{\sc ii}\,$\lambda 1640$
emission and on Al\,{\sc iii}\,$\lambda 1853-1862$ and Mg\,{\sc
ii}\,$\lambda 2800$ absorption features 
(Fig.~\ref{fig2}f). There are no outstanding
spectral characteristics marking this object as the true optical
counterpart.  K3 is a clear absorption-line galaxy at $z=1.02$, which
is also detected in the deep {\it ISO} {\it CAM} observations of this
field (J.-P.\ Kneib, unpublished). However, its large separation (6.3$''$)
from the submm position makes it a less likely counterpart.

\noindent{\bf SMM\,J00265+1710:} M6/M10 are a pair of emission-line
galaxies at $z=0.21$ (Fig.~\ref{fig2}g), 
foreground to the cluster, which appear to be
interacting. We were unable to obtain a redshift for the other
candidate counterpart, M7, because of its close proximity to M6. The
apparent interaction of the M6/M10 pair may be sufficient justification
to suggest that this is the optical counterpart to the submm
source.  However, the spectrum of neither galaxy is particularly unusual,
and we note that the galaxies are not associated with strong 15-$\mu$m emission
in the deep {\it ISO} {\it CAM} image of this cluster (J.-P.\ Kneib,
unpublished).  If the SCUBA source is associated with these galaxies
then the absence of 15-$\mu$m emission is unusual given their low
redshift. For this reason we caution that, while attractive, the
identification of the submm emission with M6/M10 may be incorrect.

\noindent{\bf SMM\,J22472$-$0206:} P2 is the only likely visible
optical counterpart to the submm detection. Our best estimate is that
it is a star-forming galaxy at $z=2.11$, based on what appears to be
Ly\,$\alpha$ emission and the Lyman break (Fig.~\ref{fig2}h). 
However, there is no
absorption-line confirmation, and the ultraviolet end of the spectrum
is obtained from an integration of only 1\,hr.  This redshift
identification is therefore uncertain.

\noindent{\bf SMM\,J00266+1710:} M3 is an emission-line galaxy at
$z=0.94$ with [O\,{\sc ii}]\,$\lambda 3727$ and
[O\,{\sc iii}]\,$\lambda 5007$ emission lines.
There is no sign of Mg\,{\sc ii}\,$\lambda 2800$ absorption or emission
in this galaxy, but we believe that the redshift identification is
reliable. In Fig.~\ref{fig3} we show the 2D spectral image for this galaxy with
tick marks at the positions of [O\,{\sc ii}]\,$\lambda 3727$, H$\beta$,
and [O\,{\sc iii}]\,$\lambda 5007$. 
M3 is the only visible optical counterpart in the vicinity of the submm 
position.

\noindent{\bf SMM\,J14014+0252:} The core of the cluster targeted in
this field, Abell\,1835, contains a massive cooling flow centered on the
$z=0.25$ cD galaxy close to the submm source. The strong line-emission and
other spectral features of massive star formation detected in this
galaxy (\markcite{allen95}Allen 1995) suggest that it is the source of 
the submm emission. \markcite{edge99}Edge et al.\ (1999) identify a
cool dust contribution in the observed submm spectral energy 
distribution of this galaxy and suggest that it arises from dust heated 
by star formation.

\noindent{\bf SMM\,J00267+1709:} There is no visible optical counterpart 
to this source on the deep {\it HST} F814W exposure,
placing a limit of $I\geq 26$ on the apparent magnitude of any counterpart.

\noindent{\bf SMM\,J04433+0210:} This is the faintest submm source to
make it into the catalog.  Moreover, this source lies off the original
{\it HST} {\it WFPC2} exposure of this cluster used by 
\markcite{smail98}Smail et al.\ (1998) to identify candidate counterparts.  
A recent deep Keck $I$-band image indicates that an $I=24.2$ object lies 
within 1.2$''$ of the submm position.  There are no brighter sources 
within 6$''$ of this position.  The probability that this faint object is the
counterpart of the submm source is $P=0.04$ (calculated in the same
manner as in \markcite{smail98}Smail et al.\ 1998).  
This suggests that the faint optical
source may be associated with the submm emission.  However, as this
candidate was only recently acquired and is extremely faint, no
spectroscopic observations have yet been undertaken.

%
% TABLE 2
%
\begin{deluxetable}{lccccrc}
\tablewidth{500pt}
\scriptsize
\tablenum{2}
\label{table-2}
\tablecaption{Redshift catalog for the candidate optical counterparts to
the submm sample\label{tab2}}
\tablehead{
\colhead{Submm} & \colhead{850\,$\mu$m} &
\colhead{Candidate Optical} &
\colhead{$I$} &
\colhead{Counterpart} &
\colhead{Cluster} &
\colhead{Lensing} \\
\colhead{Source} & \colhead{Flux (mJy)}
& \colhead{Counterparts}
& \colhead{Mag}
& \colhead{Redshifts}
& \colhead{Name/Redshift}
& \colhead{Amplification}
}
\startdata
\multispan7{$4\sigma$ Detections  \hfil}\\
\noalign{\medskip}
SMM\,J02399$-$0136 & 25.4 & L1/L2 & 20.41 & 2.80/2.80$^a$ & A\,370/0.37        & 2.4\\
SMM\,J00266+1708 & 18.6  & M2    & 22.38 & 1.23        & Cl\,0024+16/0.39   & 1.6\\
                 &       & M11   & 22.56 & 1.06        &                    & 1.5\\
                 &       & M1   & 21.99 & 0.39        &                     & 1\\
                 &       & M8   & 21.99 & 0.44        &                     & 1.05\\
SMM\,J09429+4658 & 17.2  & H1   & 19.24 & 0.33$^b$    & Cl\,0939+47/0.40   & 1\\
                 &       & H4   & 22.15 & star        &                    & \nodata \\
SMM\,J14009+0252 & 14.5  & J3   & 23.55 & 1.5--2.5    & A\,1835/0.25       & $<1.7$\\
                 &       & J4   & 21.62 &             &                    & \nodata \\
SMM\,J14011+0252 & 12.3  & J1/J2 & 20.32 & 2.55/2.55   & A\,1835/0.25       & 2.7 \\
SMM\,J02399$-$0134 & 11.0 & L3    & 20.52 & 1.06       & A\,370/0.37        & 2.5 \\
                 &       & L4   & 21.50 & 0.42         &                    &  1.1\\
                 &       & L5   & 19.00 & 0.37$^c$       &                    & 1 \\
SMM\,J22471$-$0206 & 9.2 & P4   & 21.72 & 1.16        & Cl\,2244$-$02/0.33 & 1.9\\
                 &       & P1   & 22.92 & $\simeq 2?$ &                    & 2.3\\
                 &       & P3   & 23.17 &             &                    & \nodata \\
SMM\,J21536+1741 & 9.1   & cD   & 15.94 & 0.23        & A\,2390/0.23       & 1\\
SMM\,J02400$-$0134 & 7.6 & \nodata & $>26$ & $z>4?$  & A\,370/0.37        & $>1.9$\\
SMM\,J04431+0210 & 7.2   & N1   & 18.42 & 0.18$^d$   & MS\,0440+02/0.19   & 1\\
\\
\multispan7{$3\sigma$ Detections \hfil} \\
\noalign{\medskip}
SMM\,J21536+1742 & 6.7    & K2 & 24.69 & 1.60?             & A\,2390/0.23     & 1.9 \\
                 &        & K3 & 21.36 & 1.02              &                  & 1.7\\
SMM\,J00265+1710 & 6.1    & M6/M10 & 20.53 & 0.21/0.21$^e$ & Cl\,0024+16/0.39 & 1\\
                 &        & M7 & 21.19 &                   &                  & \\
SMM\,J22472$-$0206 & 6.1  & P2 & 24.05 & 2.11?             & Cl\,2244$-$02/0.33 & 2.2\\
SMM\,J00266+1710 & 5.9    & M3 & 23.08 & 0.94              & Cl\,0024+16/0.39 & 3.6\\
SMM\,J14010+0252 & 5.4    & cD & 15.50 & 0.25              & A\,1835/0.25     & 1\\
SMM\,J00267+1709 & 5.0    & \nodata & $>25$ & $z>4?$     & Cl\,0024+16/0.39 & $>2.2$\\
SMM\,J04433+0210 & 4.5    & N5 & 24.3 &                   & MS\,0440+02/0.19 &\nodata \\
\enddata
\tablecomments{Table entries are ordered by apparent 850-$\mu$m flux
density. The most likely optical counterpart is listed first in column 3.
The lensing amplifications were determined from the detailed mass models of
the clusters. These were constructed using the LENSTOOL ray-tracing code
on lensed features identified in high resolution optical images
(Kneib et al.\ 1993).
Uncertain redshifts are followed by a question mark.}
\tablerefs{
$a$ from Ivison et al.\ (1998) --
$b$ from Dressler et al.\ (1999) --
$c$ from Mellier et al.\ (1988) --
$d$ see also Gioia et al.\ (1998) --
$e$ see also Dressler \& Gunn (1992).
}
\end{deluxetable}

Since many of our most probable counterparts do not show any unusual
spectral features, it is not always certain that we have identified
the true optical counterpart to each submm source. 
Two of the submm detections in the
sample have no visible optical counterparts in very deep imaging, and
it is possible that the true counterparts to some other sources in the
sample are similarly optically faint. In principal such sources could 
either be at 
very high redshift or be so highly obscured that they are emitting their 
energy almost entirely in the submm (\markcite{dey99}Dey et al.\ 1999).

\section{Discussion and Conclusions}\label{disc}

We have carried out a spectroscopic study of the candidate optical
counterparts for 14 of the 17 submm sources 
(excluding the two blank fields and 
the unobserved counterpart to SMM\,J04433+0210)
from a gravitationally-lensed submm-selected
survey to determine redshifts and crucial
spectral information for a representative sample of the submm source
population. This information is essential for understanding the
evolution of obscured star formation in the Universe and the
contribution of AGN to the FIR background.

Our survey has produced firm identifications for four of the
submm sources: two central cD galaxies in the lensing clusters
(SMM\,J21536+1741 and SMM\,J14014+252), an interacting pair of galaxies
at $z=2.80$ (SMM\,J02399$-$0136 L1/L2), one of which hosts a Seyfert-2
nucleus (L1), and a further pair of galaxies at $z=2.55$
(SMM\,J14011+0252 J1/J2) that show starburst features. It has also
produced reliable identifications
for a further two weak AGN sources at $z=1.06$ (SMM\,J02399$-$0134 L3)
and $z=1.16$ (SMM\,J22471$-$0206 P4). 
In the following we refer to the four non-cluster sources listed above
as our reliable sample.

%
% Figure 4
%
\hbox{~}\smallskip
\figurenum{4}
\centerline{\psfig{figure=barger.fig4.ps,width=3.1in}}
%\figcaption[barger.fig4.ps]{
\noindent{\scriptsize\addtolength{\baselineskip}{-3pt}
{\sc Fig.}~4.---
The top panel shows the redshift distribution for the most likely optical
counterparts to the submm detections in our spectroscopic survey.
The cross-hatched histogram represents the reliable sample
of four non-cluster counterparts (L1/L2, J1/J2, L3, and P4)
and the open histogram represents cluster contamination (N1 and the two cDs).
The redshift for J3 is uncertain and only expected
to be in the range $z=1.5$--2.5; thus, it has been plotted at $z=2$.
The three models discussed in the text, including Model E from Guiderdoni
et al.\ (1998) (dotted), the original Gaussian
model from Blain et al.\ (1999a) (dashed), and the revised Gaussian model
(solid curve; see text) are superimposed on the
redshift distribution in panel one. The lower two panels give the results
published from the SCUBA survey of the Hubble Deep Field by
Hughes et al.\ (1998) and from the on-going survey of the CFRS fields
by Lilly et al.\ (1999). The shaded histograms in the bottom two panels
represent the spectroscopic identifications and the dotted histograms
the photometrically-derived limits.
The number of blank fields in our survey and in the CFRS survey are
indicated by floating boxes at nominally high redshift.
\label{fig4}

\addtolength{\baselineskip}{3pt}}
\hbox{~}\smallskip

The remaining eight sources have candidate counterparts with redshifts ranging
from $z=0.18$--2.11, but they lack remarkable spectral characteristics 
to clearly identify them as the true counterparts.
The lower redshift ($z\ll 1$) systems include a
pair of interacting galaxies at $z=0.21$ (SMM\,J00265+1710 M6/M10) and
two bright spiral galaxies at $z=0.18$ (SMM\,J04431+0210 N1; cluster member) 
and $z=0.33$ (SMM\,J09429+4658 H1).

In the top panel of Fig.~\ref{fig4}
we show our differential redshift distribution for the most likely optical 
counterparts to our sample of lensed submm detections
(excluding the unobserved counterpart to SMM\,J04433+0210). 
The redshift distribution suggests that the majority of the sources are
likely to lie at redshifts $z=1$--3. The lens amplification is not expected
to significantly distort the $N(z)$ distribution given the high
redshifts of the bulk of the counterparts.
In the bottom two panels of Fig.~\ref{fig4} we show for comparison
the redshift distributions for the on-going Canada France Redshift Survey
(CFRS, \markcite{lilly99}Lilly et al.\ 1999) and for the
Hubble Deep Field (HDF, \markcite{hughes98}Hughes et al.\ 1998).

%
% Figure 5
%
\figurenum{5}
\hbox{~}\smallskip\centerline{\psfig{figure=barger.fig5.ps,width=3.1in}}
%\figcaption[barger.fig5.ps]{

\noindent{\scriptsize\addtolength{\baselineskip}{-3pt}
{\sc Fig.}~5.---
Cumulative redshift distribution (thick solid line), excluding
cluster contamination, our two blank fields, and the unobserved source. 
Cumulative distributions for the three models
discussed in Fig.~4 and in the text are also shown:
Guiderdoni et al.\ (1998) Model E (dotted), Blain et al.\ (1999a)
Gaussian model (dashed), and revised Gaussian model (solid; see text).
\label{fig5}

\addtolength{\baselineskip}{3pt}}
\hbox{~}\smallskip

We next compare our observed redshift distribution with a few representative 
models formulated to fit the FIR background and various infrared 
counts using simple analytic descriptions of the evolution of luminous 
FIR galaxies. We superimpose on our redshift distribution in 
Fig.~\ref{fig4} the 
Gaussian model (dashed line) from \markcite{blain99a}Blain et al.\ (1999a) 
and Model~E (dotted line) from \markcite{guider98}Guiderdoni et al.\ (1998).
These models are composed of normal star forming galaxies with an additional 
population of highly obscured galaxies.

In Fig.~\ref{fig5} we plot the cumulative distributions for our non-cluster 
data (also excluding the blank sources) and the models of Fig.~\ref{fig4}.
When we perform the Kolmogorov-Smirnov (K-S) test on each model compared to
the above data set, under the assumption that we have correctly identified the 
majority of the optical counterparts, we find low probabilities of 0.012
and 0.0004 for the Gaussian model and Model~E, respectively;
thus, both of these models can be rejected for predicting a 
median redshift for the submm emission that is too high. However, when we do 
the same comparison with the reliable sample of four 
non-cluster counterparts (L1/L2, J1/J2,
L3, and P4), we find higher probabilities of 0.53 and 0.11, respectively,
neither of which are significantly inconsistent with the models.

One way to reduce the mean redshift in the Blain et al.\ model while
retaining the fits to the FIR background and counts is to
change their dust emissivity spectral index from 1.5 to 1.0. The 
850-$\mu$m flux density of galaxies at $z<2$ in the models 
is then increased as compared with galaxies at higher redshifts.
Figures~\ref{fig4} and \ref{fig5} 
show this modified model as a solid curve. The K-S test
gives probabilities of 0.15 and 0.997 when this modified 
model is compared with the full and reliable samples, respectively.

%
% Figure 6
%
\figurenum{6}
\hbox{~}\smallskip\centerline{\psfig{figure=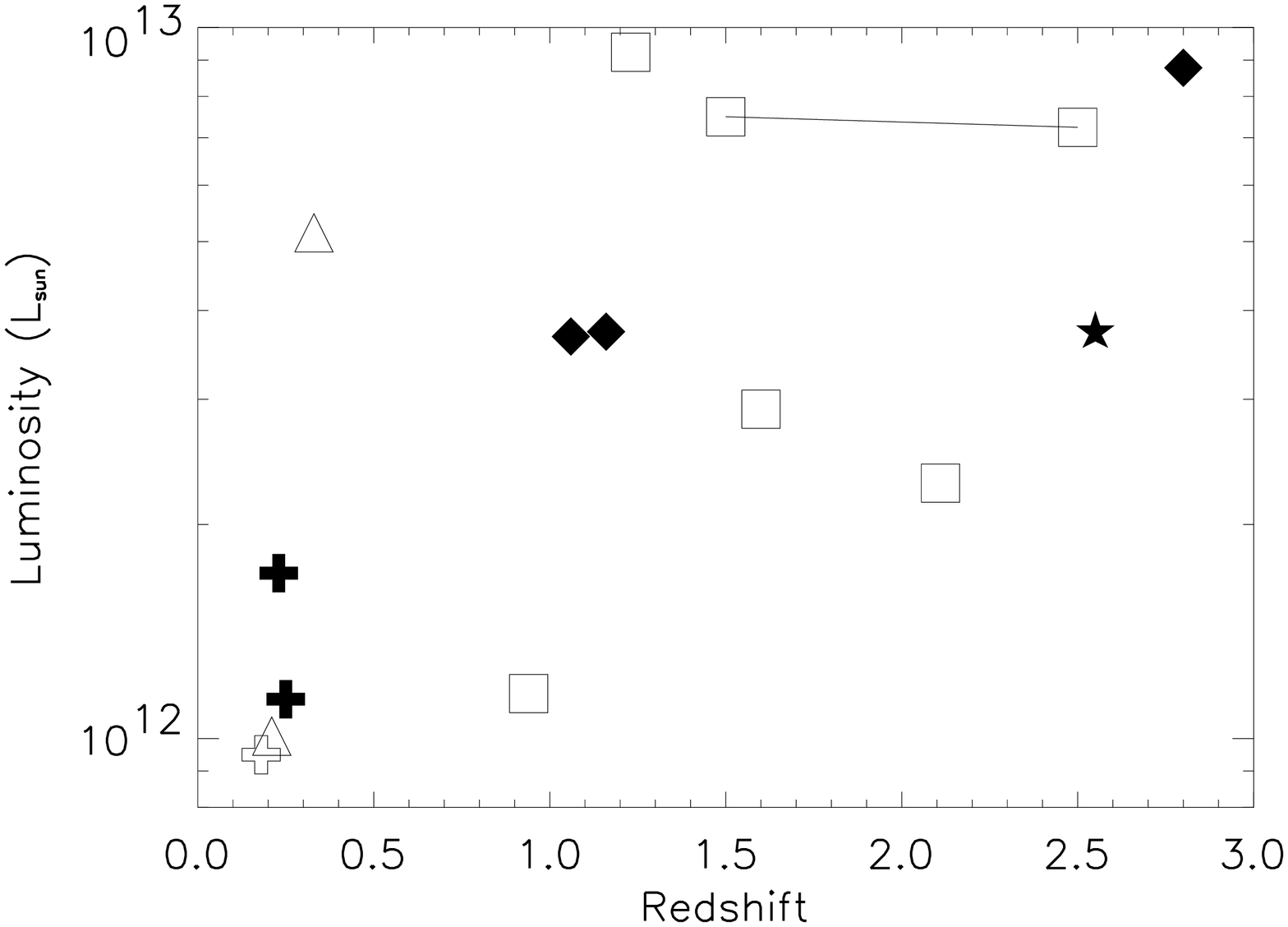,angle=90,width=3.5in
}}
%\figcaption[barger.fig6.ps]{

\noindent{\scriptsize\addtolength{\baselineskip}{-3pt}
{\sc Fig.}~6.--- 
Bolometric luminosities derived from the lensing-corrected 850-$\mu$m fluxes
assuming a dust temperature of 47\,K (Arp~220), a dust emissivity spectral
index of 1.0, and the lens amplifications listed in Table~2.
Filled symbols correspond to secure optical identifications:
the filled diamonds are objects that show AGN activity, the filled star is
our starburst pair, and the filled crosses are
the two cD galaxies. Open symbols correspond to
uncertain optical identifications: the open cross is a cluster member,
the open triangles are $z\ll 1$ field galaxies, and the open squares are
high redshift field galaxies. Note that the redshift range for J3 is
indicated with a connecting line.
\label{fig6}

\addtolength{\baselineskip}{3pt}}
\hbox{~}\smallskip

Using the redshifts determined here, accurate cluster lens models, and
a dust emissivity spectral index of 1.0, 
we can convert the apparent fluxes for the sources into
intrinsic bolometric luminosities assuming the Arp~220 dust temperature of
47\,K (\markcite{klaas97}Klaas et al.\ 1997) and a $q_0=0.5$ cosmology with 
$H_0=50\ {\rm km}\ {\rm s}^{-1}\ {\rm Mpc}^{-1}$. These luminosities are
shown in Fig.~\ref{fig6} versus redshift and are characteristic of 
ultraluminous infrared galaxies (ULIGs).
Since submillimeter observations of nearly all luminous infrared galaxies
have been reasonably fit by single temperature dust models 
with $T=30-50$\ K (\markcite{sanders96}Sanders \& Mirabel 1996),
the temperature dependence of the luminosity introduces a factor of only
about 0.2--1.2 uncertainty in the above L$_{bol}$ numbers.

A very large fraction of local ULIGs
show signatures of interactions and mergers. A relatively
high fraction of disturbed or interacting counterparts were also uncovered in
the optical identifications of the faint submm sample analyzed here
(\markcite{smail98}Smail et al.\ 1998).
Our spectroscopy has shown that although a small number
of these interacting systems are simply projection effects
(e.g.\ M1/M2), the majority are real; thus, we can
infer that interactions remain an important triggering mechanism for
ultraluminous activity in the distant Universe. This conclusion
supports the use of nearby ULIGs as templates to understand the
evolution of these more distant systems. 

In Fig.~\ref{fig7} we show a more quantitative comparison of the properties 
of the local and distant ULIGs. We compare the observed optical to submm
flux ratio versus redshift to that of the redshifted 
archetypical local ULIG Arp\,220 (solid curve). Interestingly,
the four sources with firm identifications (L1/L2, J1/J2, A1835-cD,
A2390-cD) all lie above the Arp\,220
distribution in Fig.~\ref{fig7}. However, a comparison of the
properties of the reliable non-cluster counterparts (L1/L2, J1/J2,
L3, and P4) indicates that there is considerable scatter in the
S$_I$/S$_{850}$ ratio. For the non-cluster galaxies detected in the
$I$-band, we find a median ratio of
$\sim 0.3$\,(S$_I$/S$_{850}$)$_{\rm Arp 220}$ with a dispersion of 
around a factor of four. 
The majority of our submm sources therefore appear to be emitting a
slightly lower fraction of their luminosity in the optical relative to
the submm as compared to that expected from an Arp\,220-like source
at their redshifts. We note that the blank sources would have a
S$_I$/S$_{850}$ ratio of around $1-2\times 10^{-5}$; thus, if they are
similar in their properties to our other sources, they would lie at
redshifts $z>5$. Alternatively, they could be more obscured systems at
lower redshifts.

%
% Figure 7
%
\figurenum{7}
\hbox{~}\smallskip\centerline{\psfig{figure=barger.fig7.ps,width=3.1in}}
%\figcaption[barger.fig7.ps]{

\noindent{\scriptsize\addtolength{\baselineskip}{-3pt}
{\sc Fig.}~7.--- 
Ratio of $I$-band flux to 850-$\mu$m flux versus redshift.
The line shows the relationship
versus redshift expected for Arp\,220. The
galaxies show a large scatter in their relative S$_I$/S$_{850}$ ratios
with the majority being slightly fainter in the optical relative to
their submm emission when compared to Arp\,220 put at the same redshift.
\label{fig7}

\addtolength{\baselineskip}{3pt}}
\hbox{~}\smallskip

In terms of their bolometric luminosities,
optical-to-FIR ratios, and morphologies, the galaxies selected in the
submm in the distant Universe have very similar properties to local ULIGs.
As regards the dominant energy source for the emission seen in the
FIR background, i.e.\ AGN or starburst, we find that at
least three of the fourteen sources surveyed have counterparts with
spectral features indicative of AGN activity.  These objects may be
part of the obscured AGN population predicted to be a major contributor
to the X-ray background at energies $>2$\,keV (\markcite{madau94}Madau,
Ghisellini, \& Fabian 1994; \markcite{comastri95}Comastri et al.\ 1995;
\markcite{fabian98}Fabian et al.\ 1998; \markcite{gunn99}Gunn \& Shanks
1999; \markcite{omar99}Almaini, Lawrence \& Boyle 1999).  Deep radio,
NIR, MIR, and high-resolution X-ray data of fields observed with SCUBA
should provide us with improved position estimates and more information
on the nature of the sources, making the task of following up the submm
detections easier in the future.

The determination of the redshift distribution of submm-selected
galaxies is an important goal because it allows plausible models of
the evolution of the volume emissivity of dust with redshift to be
constrained more strongly than by the background and counts
data alone. The first results
on the redshift distribution presented here suggest that we can rule
out models for galaxy and AGN evolution in which the bulk of the dust
emission in the Universe occurred at redshifts either below $z\sim 1$
or substantially higher than $z\sim 3$. This is in accord with
conclusions reached by recent modelling of the FIR background 
(\markcite{dwek98}Dwek et al.\ 1998). Consequently,
activity associated with these luminous dust-obscured sources occurs at
a similar epoch to that seen for unobscured star-forming galaxies
(\markcite{madau98}Madau, Pozzetti, \& Dickinson 1998;
\markcite{steidel99}Steidel et al.\ 1999) and AGN
(\markcite{boyle98}Boyle \& Terlevich 1998). Our spectroscopic survey 
of the probable optical counterparts to submm sources is the first 
step towards the full reconstruction of the emission history
of dusty galaxies in the Universe, whose contributions to star formation 
are now known to be at least as important at high redshift as the 
contributions from optically observed galaxies.

\smallskip
\acknowledgments
We thank an anonymous referee for useful comments.
AJB acknowledges support from NASA through contract number P423274 from the
University of Arizona, under NASA grant NAG5-3042. IRS acknowledges support
from the Royal Society. RJI and AWB acknowledge support from PPARC.
JPK acknowledges support from CNRS, a CNES/INSU grant and a EU-TMR grant.

%\newpage

\end{document}